# Topolectrical space-time circuits

Weixuan Zhang[*], Wenhui Cao[*], Long Qian[*], Hao Yuan and Xiangdong Zhang[#]

*Key Laboratory of advanced optoelectronic quantum architecture and measurements of Ministry of Education, Beijing Key Laboratory of Nanophotonics & Ultrafine Optoelectronic Systems, School of Physics, Beijing Institute of Technology, 100081, Beijing, China*

[*]*These authors contributed equally to this work.*

[#]*To whom correspondence should be addressed. E-mail: zhangxd@bit.edu.cn;*

## Abstract

**Topolectrical circuits have emerged as a pivotal platform for realizing static topological states that are challenging to construct in other systems, facilitating the design of robust circuit devices. In addition to spatial dimensionality, synergistic engineering of both temporal and spatial degrees in circuit networks holds tremendous potential across diverse technologies, such as wireless communications, non-reciprocal electronics and dynamic signal controls with exotic space-time topology. However, the realization of space-time modulated circuit networks is still lacking due to the necessity for flexible modulation of node connections in both spatial and temporal domains. Here, we propose a new class of topolectrical circuits, referred to as topolectrical space-time circuits, to bridge this gap. By designing and applying a novel time-varying circuit element controlled by external voltages, we can construct circuit networks exhibiting discrete space-time translational symmetries in any dimensionality, where the circuit dynamical equation is in the same form with time-dependent Schrodinger equation. Through the implementation of topolectrical space-time circuits, three distinct types of topological space-time crystals are experimentally demonstrated, including the (1+1)-dimensional topological space-time crystal with midgap edge modes, (2+1)-dimensional topological space-time crystal with chiral edge states, and (3+1)-dimensional Weyl space-time semimetals. Our work establishes a solid foundation for the exploration of intricate space-time topological phenomena and holds potential applications in the field of dynamically manipulating electronic signals with unique space-time topology.**

Time-periodic driving is a powerful tool for controlling both quantum and classical systems, offering the potential to facilitate on-demand dynamical manipulation of material properties and transcend certain physical limitations inherent in static systems. In the past few years, the significant progress has been made in elucidating the unique properties of representative time-varying systems, such as the discrete time crystal [1, 2], topological pumping [3], photonic time crystals [4-6], Floquet topological matters [7-16] and others [17-22]. Recently, there has been a proposal to extend the separated temporal and spatial periodicities of crystals to encompass the general (1+D)-dimensional space-time translation symmetry [23]. The system with discrete space-time translation symmetries corresponds to the space-time crystal, including both spatial crystals and Floquet crystals as special cases. Interestingly, it has been demonstrated theoretically that diverse types of non-equilibrium topological phases can exist in space-time crystals without spatial periodicity [24-27]. For example, a recent theoretical study has suggested a scheme for the realization of (1+1)- and (2+1)-dimensional topological space-time crystals protected by $Z_2$ and $Z$ topological invariants, where time-dependent drives resembling traveling waves are introduced [24]. The constructed topological space-time crystals exclusively involve a single orbital, distinguishing them from previously discovered static or Floquet topological phases with crystalline structures that require at least two orbitals. Despite the novel properties and potential applications of Floquet topological states with intertwined space-time symmetries, significant challenges persist in their experimental implementation due to the intricate requirement of simultaneously modulating systematic parameters for both spatial and temporal degrees. Thus, exploring new strategies for addressing these challenges and experimentally engineering space-time topologies in arbitrary dimensions remains to be investigated.

On the other hand, building upon the consistency of circuit networks and tight-binding lattice models, topolectrical circuits have emerged as a powerful platform for investigating topological physics [28-67]. In contrast to quantum materials and classical-wave metamaterials, electric circuits can exhibit nonlocal couplings independent of spatial dimensions and node distances, enabling reliable experimental explorations of high-dimensional [40-42], non-Euclidean [43-46] and non-Abelian [47-49] topological states that are hard to be constructed in other systems. Moreover, due to the availability of versatile circuit elements such as operational amplifiers, memristive elements and multipliers, circuits offer flexible control over various physical properties that are either absent in natural materials or challenging to engineer in other artificial structures. This enables the realization of diverse topological phases with exotic non-Hermitian [50-59] and nonlinear [60-64] properties. In addition to the spatial dimensionality, introducing the temporal modulation into circuit networks can promote lots of intriguing phenomena that could be used in a range of technologies, such as 5G wireless communications [14, 65], non-reciprocal electronics [18], synthetic dimensions [66] and topological pumping [67]. However, the simultaneous engineering of temporal and spatial degrees in electric circuits has not yet been reported, leaving the interplay between space-time symmetries and topological physics largely unexplored in topolectrical circuits—particularly within experimental investigations. It is worth noting that space-time-modulated topolectrical circuits have the capability to manipulate both the momentum and frequency of electronic signals, offering space-time reconfigurable functionalities that surpass those of static topolectrical circuits. Therefore, how to incorporate topological space-time manipulations into electric circuit networks is a crucial issue to be resolved.

In this work, we firstly extend the concept of topolectrical circuits to the realm of topolectrical space-time circuits, and experimentally implement three types of Floquet topological states with distinct space-time translational symmetries. Specifically, the midgap edge modes protected by the generalized particle-hole symmetry have been directly observed in (1+1)-dimensional topolectrical space-time circuits. Moreover, by constructing two-dimensional circuit networks with (2+1)-dimensional space-time translation symmetries, we experimentally observe the chiral propagation of topological space-time edge states through measuring the voltage evolution. Finally, we also theoretically propose the (3+1)-dimensional space-time Weyl semimetal, and experimentally fulfil topological surface states induced by space-time Weyl points by three-dimensional topolectrical space-time circuits. Our work establishes a solid foundation for investigating intricate space-time topological phenomena and holds immense potential for future experimental simulations involving diverse time-modulated Hamiltonians with exceptional dynamical properties.

**Results**

**(1+1)-Dimensional topolectrical space-time circuits with mid-gap edge modes.** We consider a time-varying one-dimensional lattice model along the $x$-direction, where the coupling strength between two adjacent sites is dependent on both spatial position and time. The tight-binding Hamiltonian of our model is written as

$$H_{1D}(x,t) = \sum_{x\in[1,L]}\{i[J_0 + J(x,t)]a_x^\dagger a_{x+1} + H.c.\}, \quad (1)$$

where $a_x^\dagger$ ($a_x$) is the creation (annihilation) operator at position $x$ with lattice length being $L$. $J_0$ is the time-independent hopping term, and $J(x,t) = \Delta\cos[k_\delta(x + 0.5) - \Omega t]$ corresponds to the time-varying hopping term with a traveling wave profile. In this case, the lattice Hamiltonian lacks spatial periodicity at any given moment in time. Instead, the system exhibits a discrete space-time translation symmetry $H(x,t) = H(x + 1, t + k_\delta/\Omega)$, where space and time are coupled, along with an additional discrete time translation symmetry $H(x,t) = H(x, t - 2\pi/\Omega)$. Thus, our model can be regarded as a (1+1)-dimensional space-time crystal with two space-time translation vectors $T_{i=1,2} = (s_i, \tau_i)$ being $T_1 = (0, -2\pi/\Omega)$ and $T_2 = (1, k_\delta/\Omega)$. The corresponding reciprocal vectors $K_{i=1,2} = (G_i, \omega_i)$ in the momentum-energy space are $K_1 = (k_\delta, \Omega)$ and $K_2 = (2\pi, 0)$ with $s_i \cdot G_j - \tau_i\omega_j = 2\pi\delta_{ij}$. It is worth noting that our model (as well as the subsequent (2+1)- and (3+1)-dimensional cases) represents a specific type of (D+1)-dimensional space-time crystals, where the system exhibits D discrete space-time translation symmetries and a discrete time translation symmetry. Importantly, this discrete time translation symmetry ensures that only one nonzero $\omega_i$ exists in the reciprocal vectors $(G_i, \omega_i)$, being a necessary condition for the existence of quasi-energy band gaps [24]. Because, for any solution with $(k, \omega)$, if there are two nonzero $\omega_i$ labeled by $\omega_1$ and $\omega_2$ in the reciprocal vectors, all states of the form $(k + nG_1 + mG_2, \omega + n\omega_1 + m\omega_2)$ with $n, m \in \mathbb{Z}$ will also be valid solutions. Since $\omega_1$ and $\omega_2$ are incommensurate, $n\omega_1 + m\omega_2$ can approach zero for large values of $|n|$ and $|m|$, leading to a gapless spectrum of quasi-energy.

By applying the generalized Floquet-Bloch theorem [24], the eigenstate labeled by quasi-momentum $k_x$ and quasi-energy $\varepsilon$ of the 1D space-time crystal is in the form of $\psi_{k_x,\varepsilon}(x,t) = e^{i(k_x x - \varepsilon t)}\mu_{k_x,\varepsilon}(x,t)$ with $\mu_{k_x,\varepsilon}(x,t)$ satisfying the same discrete space-time translation symmetries to the crystal Hamiltonian. In this case, similar to the conventional Floquet-Bloch system, the (1+1)-dimensional space-time crystal can also be described by an energy-domain-enlarged Floquet Hamiltonian in quasi-momentum space. It is noted that both the enlarged Hamiltonian with respect to the quasi-energy $\varepsilon = 0.5\Omega$ labeled by $H_{\varepsilon=0.5\Omega}(k_x)$ and the low-energy effective Hamiltonian $H_{eff}(k_x)$ of two neighboring Floquet sectors around $\varepsilon = 0.5\Omega$ satisfy the generalized particle-hole symmetry as $\hat{C}H_{\varepsilon=0.5\Omega}(k_x)\hat{C}^{-1} = -H_{\varepsilon=0.5\Omega}(-k_\delta - k_x)^*$ and $\hat{C}_e H_{eff}(k_x)\hat{C}_e^{-1} = -H_{eff}(-k_\delta - k_x)^*$ (See Supplementary Note 1 for details). It should be noted that the enlarged Hamiltonian $H_{\varepsilon=0.5\Omega}(k)$ is simply obtained by a trivial energy shift from the original Floquet Hamiltonian $H(k)$, specifically as $H_{0.5\Omega}(k) = H(k) + 0.5\Omega I$. This energy shift does not affect the system's symmetry, and thus, the classification of the space-time Hamiltonian is independent of the reference energy. Hence, according to Altland-Zirnbauer topological classification with symmetries [68], the proposed (1+1)-dimensional space-time crystal with particle-hole symmetry belongs to the class $D$ described by $Z_2$ topological invariants, and is expected to support midgap topological states at $\varepsilon = (0.5 + n)\Omega$ with $n = 0, \pm 1, ...$, even if the system is non-crystalline with $k_\delta$ being an irrational number or $L$ being smaller than the period with respect to $k_\delta$.

We calculate the eigenspectrum of the real-space Floquet Hamiltonian for the (1+1)-dimensional space-time crystal with open boundaries. Figure 1(a) presents the quasi-energy spectrum with $L$=31, and other parameters are taken as $J_0 = 0.5\Omega$, $\Delta = 0.5\Omega$, and $k_\delta = 0.81\pi$. The color map represents the quantity of $S(\varepsilon) = \sum_{x \in edge} |\phi_\varepsilon(x)|^2 / \sum_{x=[1,L]} |\phi_\varepsilon(x)|^2$, which characterizes the strength of the edge localization for eigenmode $\phi_\varepsilon(x)$ with quasi-energy $\varepsilon$. It is clearly shown that the mid-gap topological states with large-valued $S(\varepsilon)$ exist at $\varepsilon = \pm 0.5\Omega$. The corresponding spatial profile is depicted in Fig. 1(b), where the strong boundary localization manifests the presence of topological edge states. To further illustrate the phase diagram of (1+1) topological space-time crystals, we calculate the quasi-energy spectrum of the finite chain as a function of $k_\delta$ at fixed $J_0 = 0.5\Omega$ and $\Delta = 0.5\Omega$, as shown in Fig. 1(c). The results show that the midgap space-time edge states emerge within the range of $k_\delta \in [0.59\pi, 1.41\pi]$, indicating that topological phase transitions of (1+1)-dimensional space-time crystals occur at $k_\delta = 0.59\pi$ and $1.41\pi$.

It is noted that the tight-binding lattice model of the (1+1)-dimensional topological space-time crystal only possesses one orbit, which is different to previous two-orbital minimal models for static and Floquet counterparts. The one-orbit nature of the (1+1)-dimensional topological space-time crystal arises from the unique structure of its energy bands in the enlarged Hamiltonian within the frequency domain. Specifically, across different Floquet sectors, the diagonal energy bands of $H_{\varepsilon=0.5\Omega}(k_x)$ not only exhibit energy shifts, but also show momentum differences in multiples of $k_\delta$. As a result, two diagonal energy bands of the same orbit from adjacent Floquet sectors can intersect at specific $k$-points. When these bands are effectively coupled through the time modulation, a topological band gap can open. In contrast, for conventional Floquet systems with separated space and time translational symmetry, i.e. $k_\delta = 0$, diagonal energy bands from adjacent Floquet sectors do not intersect, preventing the formation of a topological band gap. To open a topological band gap in conventional Floquet-Bloch systems, at least two orbits are required. In this case, bands from different orbits in adjacent Floquet sectors can intersect, allowing the off-diagonal block to couple these intersecting bands and open a topological band gap. The physical origin of one-orbit (1+1)-dimensional topological space-time crystals can be extended to one-orbit ($D$+1)-dimensional topological space-time crystals.

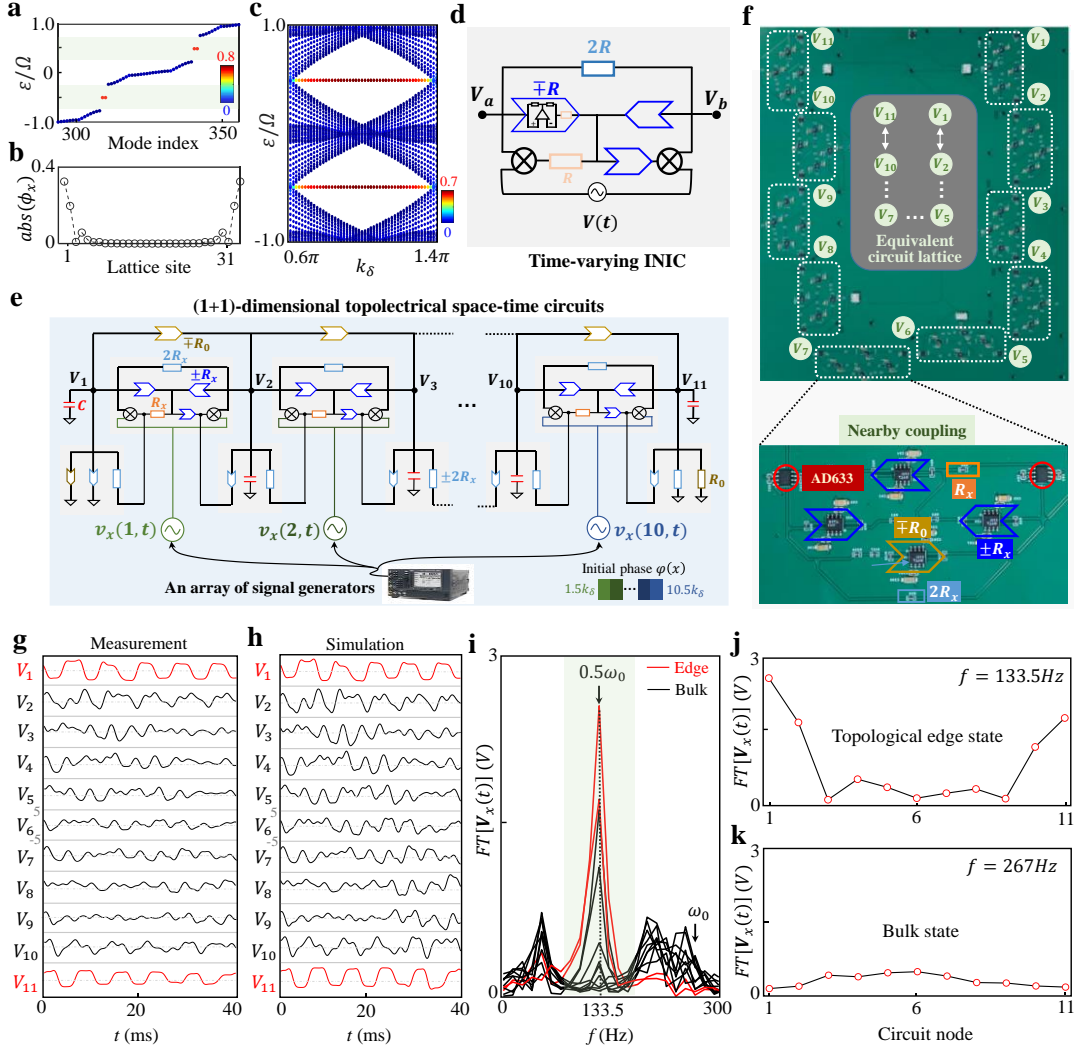

**Figure 1. Theoretical and experimental results of (1+1)-dimensional topolectrical space-time circuits.** (a). Numerical results of the quasi-energy spectrum for the (1+1)-dimensional topolectrical space-time crystal with $J_0 = 0.5\Omega$, $\Delta = 0.5\Omega$, $k_\delta = 0.81\pi$, and $L$=31. The colormap quantifies the boundary localization of all eigenmodes. (b). The spatial profile of a midgap space-time topological boundary state at $\varepsilon = 0.5\Omega$. (c). The quasi-energy spectra of the finite chains as a function of $k_\delta$. Other parameters are the same to that used in (a). (d). The schematic diagram for the implementation of a time-varying INIC connecting two adjacent circuit nodes, where the external voltage $V(t)$ is used to manipulate the detailed time behaviour. (e). The scheme of the (1+1)-dimensional topolectrical space-time circuit. The position-dependent initial phase $\varphi(x)$ of the external voltage $v_x(x,t)$ is realized by an array of signal generators. (f). The photograph image of the fabricated circuit sample with $L$=11. The inset presents the enlarged view around a single node. (g)-(h). Measured and simulated voltage waveforms of all nodes in the circuit with red and black lines showing the results of edge and bulk nodes. (i). Black and red lines show FT frequency spectra of measured voltages at boundary and bulk circuit nodes. (j)-(k). Measured spatial distributions of $FT[V_x(t)]$ at $f$=267 Hz for a bulk mode and 133.5 Hz for the midgap topological boundary mode.

To experimentally realize the topological space-time crystal, modulating the intricate site-couplings in both spatial and temporal domains is required, which are difficult in either quantum or classical-wave materials. Motivated by the experimental breakthrough in realizing topological phases by circuit networks [28-67], in the following, we explore the construction of (1+1)-dimensional topological space-time crystals by topolectrical space-time circuits.

For this purpose, we start to introduce the design of a fundamental circuit element to achieve the effective space-time-modulated coupling. Fig. 1d displays the schematic diagram illustrating the implementation of a time-varying impedance converter through current inversion (INIC) connecting two circuit nodes. The corresponding Laplacian matrix is expressed as (See Supplementary Note 2 for the detailed derivation)

$$\begin{bmatrix} I_a \\ I_b \end{bmatrix} = \frac{V(t)}{20R} \begin{bmatrix} 1 & -1 \\ 1 & -1 \end{bmatrix} \begin{bmatrix} V_a \\ V_b \end{bmatrix}, \qquad (2)$$

where $V_a$ and $I_a$ ($V_b$ and $I_b$) represent the voltage and current at the circuit node labeled by $a$ ($b$). $V(t) = V_0 \cos(\omega_0 t + \varphi)$ is the external voltage being injected into two analog multipliers, where the transfer function of the multiplier is $V_{out}(t) = V_{a/b} * \frac{V(t)}{10\ V}$ with $V_{out}(t)$ being the output voltage of the multiplier. It is shown that the currents do not fulfil the reciprocity condition of $I_a = -I_b$, indicating that our designed circuit element connecting two nodes can act as a time-varying INIC controlled by the external voltage.

By utilizing this time-varying circuit element, we can design an electric circuit ($L=11$) to implement the (1+1)-dimensional space-time lattice model of Eq. (1), as depicted in Fig. 1e. The voltage $V_x$ can be mapped to the probability amplitude of the lattice site at position $x$. The space-time-invariant hopping can be easily realized by conventional INICs $\pm \frac{1}{R_0}$, and the space-time-modulated hopping is able to be fulfilled by time-varying INICs $\pm \frac{v_x(x,t)}{20R_x}$, where the external voltages $v_x(x,t) = V_0 \cos[\omega_0 t + \varphi(x)]$ across all circuit nodes possess the same driving frequency but position-dependent initial phases $\varphi(x) = k_\delta(x + 0.5)$, as presented by the color map in Fig. 1e. In the experimental implementation, the space-time coupling terms are controlled by external voltage $v_x(x,t)$ fed into one port of each multiplier for the time-varying INIC. Specifically, an array of synchronized signal generators produces external voltage signals with identical frequencies and required initial phases. Then, using voltage-controlled, high-time-resolution switches, the external signals $v_x(x,t)$ and the initial voltages at each circuit node $V_x(t=0)$ are simultaneously applied to the circuit. In this case, the external voltage of each multiplier precisely aligns with the theoretical requirement. This setup allows us to simulate discrete-time and space-time translation symmetries within the experimental circuit system (see Methods for details). Specifically, after a specific time period $t = 2\pi/\omega_0$, the coupling strengths across the entire system return to their initial values, demonstrating a periodic reset of the system. Thus, while the circuit couplings are continuously modulated, the system returns to its original configuration after each interval of $t = 2\pi/\omega_0$, signifying the discrete-time translational symmetry. Furthermore, the time-varying circuit couplings also remain invariant under a combination of the discrete space and time translations. After a time interval $t = k_\delta/\omega_0$, and shifting all nodes with $x \to x - 1$, the circuit connection returns to its initial state, confirming that the system exhibits discrete a space-time translational symmetry. Moreover, each circuit node is also grounded with a capacitor $C$ and connects to outputs of adjacent analog multipliers to ensure the time-invariant onsite potential. In this case, the voltage dynamical equation of our designed (1+1)-dimensional topolectrical space-time circuit can be written as (see Supplementary Note 3 for the detailed derivation)

$$i\frac{d}{dt}V_x = i(\frac{v_x(x,t)}{20CR_x} + \frac{1}{CR_0})V_{x+1} - i(\frac{v_x(x-1,t)}{20CR_x} + \frac{1}{CR_0})V_{x-1}. \tag{3}$$

It is shown that Eq. (3) possesses the same form with the Schrodinger equation of (1+1)-dimensional topological space-time crystal $i\frac{d}{dt}\phi(x) = i[J(x,t) + J_0]\phi(x + 1) - i[J(x,t) + J_0]\phi(x - 1)$, where the voltage $V_x$ is analogous to the wave function $\phi(x)$. Moreover, the correspondence between tight-binding parameters and circuit elements are denoted by $\frac{v_x(x,t)}{20CR_x}/\omega_0 = J(x,t)/\Omega$ and $\frac{1}{CR_0}/\omega_0 = J_0/\Omega$. In this case, quasi-energies of the topological space-time crystal are directly mapped to quasi-frequencies ($f$) of the topolectrical space-time circuit with $\frac{2\pi f}{\omega_0} = \frac{\varepsilon}{\Omega}$. Consequently, our designed electric circuit can serve as a practical platform for studying (1+1)-dimensional topological space-time crystals. It is also important to note that the parameter selection for the designed topolectrical space-time circuit is based on two primarily considerations. Firstly, the parameters must ensure a correspondence between the voltage dynamical equation of the circuit and the tight-binding Schrödinger equation of the target lattice model. This informs the choice of capacitance, resistance, position-dependent phase factor $k_\delta$, and external driving frequency, ensuring that the circuit's effective coupling parameters mirror those of the theoretical lattice model. Secondly, the selected parameters must meet the operational constraints of the active components (such as operational amplifiers and multipliers) used in the circuit, specifically their working frequency ranges and output voltage limits.

We fabricate the designed topolectrical space-time circuit with $C = 200nF$, $R_0 = 6k\Omega$, $R_x = 1.5k\Omega$, $k_\delta = 0.81\pi$ and $L$=11, as shown in Fig. 1f, where white dash blocks enclose couplings between adjacent nodes. The inset provides an enlarged view of circuit elements connected two nearby nodes, where a conventional INIC and a time-varying INIC are applied to fulfill time-invariant and time-varying couplings. Two AD633 analog multipliers are applied to construct the time-varying INIC, where the external voltage with a position-dependent input phase is injected into one input port of each multiplier. The amplitude and angular frequency for external voltages are equal to $V_0 = 5V$ and $\omega_0 = 1.67kHz$, respectively. In this case, quasi-frequencies of our designed topolectrical space-time circuit and quasi-energies (in Fig. 1a) of the topological space-time crystal satisfy the relationship of $\frac{2\pi f}{\omega_0} = \frac{\varepsilon}{\Omega}$ with $\frac{V_0}{20CR_x}/\omega_0 = \frac{\Delta}{\Omega} = 0.5$ and $\frac{1}{CR_0}/\omega_0 = \frac{J_0}{\Omega} = 0.5$. It should be noted that the tolerance of circuit elements is limited within only 1% (0.1%) for capacitors (resistors) to prevent detuning of circuit responses, and circuit parameters are set to be sufficiently large, rendering any influence from parasitic capacitances and resistances in the circuit sample negligible. Further details on the sample fabrication and experimental setup can be found in Methods.

To observe (1+1)-dimensional topological space-time boundary states, we firstly measure voltage dynamics of all boundary and bulk circuit nodes with an initial voltage of $2V$ at all nodes, as shown in Fig. 1g by red and black lines. The corresponding simulation results by *LTSpice* are presented in Fig. 1h. It is evident that voltage waveforms of all circuit nodes align well with simulations, and the small damping of experimental voltage signals can be attributed to the loss effect originating from non-ideal performances of OpAmp and multiplier. Then, the Fourier transformation (*FT*) is performed on measured voltages, and the obtained frequency spectra (labeled by $FT[V_x(t)]$) are displayed in Fig. 1i. Notably, due to the correspondence between eigen-equations of space-time crystals and circuits, each peak of *FT* frequency spectra precisely corresponds to a quasi-mode of the (1+1)-dimensional space-time crystal. In particular, the isolated frequency peak at $f = \frac{0.5\omega_0}{2\pi} = 133.5$ Hz is matched to the in-gap topological boundary state with $\varepsilon = 0.5\Omega$, while multiple frequency peaks ranging from 0 Hz to 61 Hz and from 204.23 Hz to 265.3 Hz are consistent with quasi-energies of bulk modes from $\varepsilon = 0$ to $0.23\Omega$ and $\varepsilon = 0.77\Omega$ to $\Omega$. Then, as presented in Figs. 1j-1k, we further plot spatial distributions of absolute values for $FT[V_x(t)]$ at $f$=267 Hz (a bulk mode) and 133.5 Hz (the midgap topological mode), respectively. We can see that the spatial profile of $FT[V_x(t)]$ shows a strong boundary localization at $f$=133.5 Hz, being consistent to the midgap topological boundary state. The asymmetric boundary localization around two endpoints arises from the non-crystalline nature of the space-time circuit. In addition, the voltage is extended into the bulk region when $f$=267 Hz, manifesting the profile of a trivial bulk state. These experimental results clearly demonstrate the realization of (1+1)-dimensional topological space-time boundary states by topolectrical space-time circuits. Finally, it is worth noting that the conventional two-orbit Floquet topological state with separated spatial and temporal translational symmetries can also be easily realized by setting the initial phase of external voltage as $\varphi(x) = \pi(x + 0.5)$ for our designed topolectrical space-time circuit.

**(2+1)-Dimensional topolectrical space-time circuits with chiral edge states.** In this part, we extend the (1+1)-dimensional topological space-time crystal to (2+1)-dimensional cases to explore the exotic space-time topology through (2+1)-dimensional topolectrical space-time circuits. The Hamiltonian of our considered (2+1)-dimensional topological space-time crystal is formulated as

$$H_{2D}(x,y,t) = \sum_{x,y\in[1,L]}\{i[J_0 + J_x(x,y,t)]a^\dagger_{x,y}a_{x+1,y} - i[J_0 + J_y(x,y,t)]a^\dagger_{x,y}a_{x,y+1} + H.c.\}. \quad (4)$$

Here, $J_0$ is the constant coupling strength along the *x*- and *y*-axes. The time-varying coupling terms along *x*- and *y*-axes are $J_x(x,y,t) = \Delta\cos[k_\delta^x(x + 0.5) + k_\delta^y y - \Omega t]$ and $J_y(x,y,t) = \Delta\sin[k_\delta^x x + k_\delta^y(y + 0.5) - \Omega t]$. Thus, the Hamiltonian obey the discrete space-time translation symmetries, defined as $H(x,y,t) = H(x,y,t - 2\pi/\Omega)$, $H(x,y,t) = H(x + 1, y, t + k_\delta^x/\Omega)$ and $H(x,y,t) = H(x, y + 1, t + k_\delta^y/\Omega)$. In this case, similar to the (1+1)-dimensional space-time crystal, the (2+1)-dimensional space-time crystal can also be described by an energy-enlarged Floquet Hamiltonian in the 2D quasi-momentum space. The enlarged Hamiltonian $H_{\varepsilon=0.5\Omega}(k_x, k_y)$ with respect to $\varepsilon = 0.5\Omega$ and the low-energy effective Hamiltonian $H_{eff}(k_x, k_y)$ of two neighboring Floquet sectors around $\varepsilon = 0.5\Omega$ are both in the class *A* without time-reversal, particle-hole and chiral symmetries [68], and can be characterized by an integer (Z) topological invariant (See Supplementary Note 4 for details). Therefore, the chiral topological boundary state is able to be constructed in the (2+1)-dimensional space-time crystal with open boundary conditions. Figure 2(a1) presents the calculated quasi-energy spectrum of the finite lattice model with *L*=15 and other systematic parameters are set as $J_0 = 0.3\Omega$, $\Delta = 0.5\Omega$, and $k_\delta^x = k_\delta^y = 0.81\pi$. The colormap quantifies the boundary localization $S(\varepsilon)$ of all eigenmodes. It is clearly shown that gapless topological boundary states exist within the bulk gap around $\varepsilon=\pm 0.5\Omega$, and the spatial profile of a topological state at $\varepsilon = 0.5\Omega$ is plotted in Fig. 2(b), showing the strong boundary localization. In addition, we note that the topological boundary states near $\varepsilon = 0.5\Omega$ and $-0.5\Omega$ result from the coupling of two distinct pairs of single-orbit Floquet energy bands, which differ in both their quasi-energies and quasi-momenta. To elaborate further, the calculated edge state dispersion is shown in Fig. 2(a2), where the *x*-direction uses periodic boundary condition and the *y*-direction uses open boundary condition with fifteen lattice sites. We find that there are two gapless topological edge states within each band gap, corresponding to boundary states localized at the top and bottom edges along the *y*-axis. It is worth noting that, due to the space-time translation symmetries in the system, the gapless edge states across different Floquet sectors exhibit not only an energy shift $\varepsilon \to \varepsilon + \Omega$ but also a momentum shift $k_x \to k_x + k_\delta^x$. This behavior distinguishes these topological modes from those found in conventional Floquet-Bloch topological insulators, which do not exhibit such momentum-dependent shifts. In Supplementary Note 5, we perform numerical calculations to investigate the temporal evolution of topological boundary states in the (2+1)-dimensional topological space-time crystal with different sizes. It is found that the chiral propagation of topological edge states can exist in the system with only five units along *x*- and *y*-directions. Moreover, we present the phase diagram of the (2+1)-dimensional topological space-time crystal in ($k_\delta^x$, $k_\delta^y$) space with $J_0 = 0.3\Omega$ and $\Delta = 0.5\Omega$, as shown in Fig. 2(c). In this diagram, the space-time topological phase is identified by the presence of gapless topological edge states, along with topological band gaps exhibiting non-trivial Chern numbers for the enlarged Hamiltonian in momentum space (see Supplementary Note 6 for details). Notably, the frequency-domain formulation of topological invariants, based on the truncated Floquet Hamiltonian [69], for periodically driven systems with separated spatial and temporal translational symmetries can also be applied to describe topological space-time crystals [24]. Our results reveal that the (2+1)-dimensional space-time topological states occur in the region where the absolute values of $k_\delta^x$ and $k_\delta^y$ are approaching to π, as illustrated by blue blocks. While, the central domain corresponds to the gapless region without any band gap.

Next, we turn to the design of a (2+1)-dimensional topolectrical space-time circuit to implement the finite (2+1)-dimensional topological space-time crystal with chiral edge states. Fig. 2d illustrates the schematic diagram of the (2+1)-dimensional topolectrical space-time circuit. The space-time-invariant hopping is realized by conventional INIC $\pm\frac{1}{R_0}$, and the space-time-modulated hopping along $x$- and $y$-axis can be achieved by the external voltage-controlled INICs $\pm\frac{v_x(x,y,t)}{20CR_x}$ and $\pm\frac{v_y(x,y,t)}{20CR_y}$, as shown in the right-top inset. Specifically, the external voltages injected into time-varying INICs along $x$- and $y$-axes are in the form of $v_x(x,y,t) = V_0\cos[\omega_0 t + \varphi_x(x,y)]$ and $v_y(x,y,t) = V_0\sin[\omega_0 t + \varphi_y(x,y)]$, where the site-independent driving frequency and amplitude are equal to $\omega_0 = 1.67kHz$ and $V_0 = 5V$. In addition, the site-dependent initial phases are expressed as $\varphi_x(x,y) = k_\delta^x(x+0.5) + k_\delta^y y$ and $\varphi_y(x,y) = k_\delta^x x + k_\delta^y(y+0.5)$, as displayed by the color map in Fig. 2d. The right-bottom inset displays the values of all non-reciprocal and reciprocal resistances applied in our circuit. In this case, the voltage dynamical equation of the (2+1)-dimensional space-time topolectrical circuit takes on the same form with that of the (2+1)-dimensional topological space-time crystal as $\frac{v_x(x,y,t)}{20CR_x}/\omega_0 = J_x(x,y,t)/\Omega$, $\frac{v_y(x,y,t)}{20CR_y}/\omega_0 = J_y(x,y,t)/\Omega$ and $\frac{1}{CR_0}/\omega_0 = J_0/\Omega$ (see Supplementary Note 7 for details). Fig. 2e presents the image of the fabricated circuit with white dash blocks enclosing nearby couplings between two nodes. The inset displays the enlarged view of the nearby coupling along $y$-axis. The circuit parameters are set as $C = 200nF$, $R_0 = 10k\Omega$, $R_x = R_y = 1.5k\Omega$, $k_\delta^x = k_\delta^y = 0.81\pi$ and $L=5$, where the circuit quasi-frequency is related to the quasi-energy presented in Fig. 2a as $\frac{2\pi f}{\omega_0} = \frac{\varepsilon}{\Omega}$ with $\frac{\frac{V_0}{20CR_x}}{\omega_0} = \frac{\frac{V_0}{20CR_y}}{\omega_0} = \frac{\Delta}{\Omega} = 0.5$ and $\frac{1}{CR_0}/\omega_0 = J_0/\Omega = 0.3$.

Then, we conduct experimental measurements to detect the chiral propagation of topological space-time boundary states. In this setup, initial voltages of 2V were applied to two nodes in the lower-left and upper-right corners, while all other nodes were set to zero. By exciting the circuit at these two spatially separated points, we can effectively demonstrate the chiral propagation of topological states along distinct boundary regions, with a reduced degree of bulk state excitation compared to the single-point excitation. The spatial distributions of measured $|V_{x,y}(t)|$ at $t$=0, 3.7ms, 6.3ms, and 9.4ms are plotted in Figs. 2(f1)-2(f4). It is clearly shown that both input voltages on left-bottom and right-top corner nodes primarily propagate along boundary nodes in the clockwise direction, demonstrating the chiral behavior of topological edge states. Meanwhile, small-amplitude voltages are observed in the bulk domain during the initial period. This can be attributed to the relatively weak excitation of trivial bulk states compared to the topological edge states under the initial voltage conditions. As a result, the voltage dynamics are primarily governed by the chiral boundary propagation, with only minimal interference from bulk signals.

The good consistence between temporal measurements and simulations of voltage waveforms (see Supplementary Note 8) clearly confirm the realization of (2+1)-dimensional topolectrical space-time circuits. Moreover, we note that our (2+1)-dimensional topolectrical space-time circuit can also be used to directly implement the Floquet-version of Qi-Wu-Zhang model [70] with separated spatial and temporal translational symmetries by setting initial phases of the external voltage as $\varphi_x(x,y)=\pi(x+0.5)+\pi y$ and $\varphi_y(x,y)=\pi x+\pi(y+0.5)$.

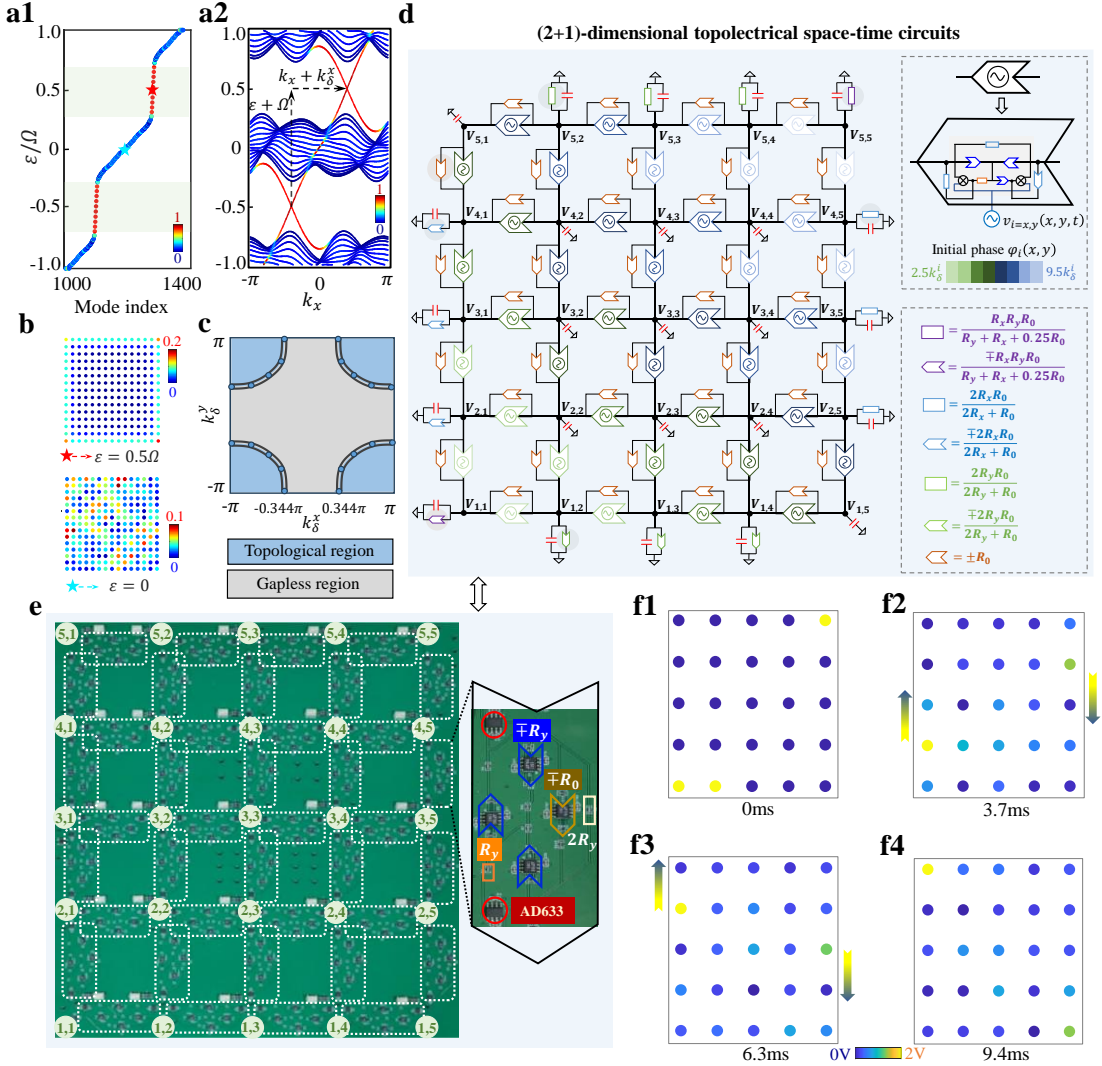

**Figure 2. Theoretical and experimental results of (2+1)-dimensional topolectrical space-time circuits.** (a1). Numerical results of the quasi-energy spectrum for the finite (2+1)-dimensional topological space-time crystal with $J_0=0.3\Omega$, $\Delta=0.5\Omega$, $k_\delta^x=0.81\pi$, $k_\delta^y=0.81\pi$, and $L$=15. The color map quantifies the boundary localization of all eigenmodes. (a2). The edge-state dispersion under mixed boundary conditions, where the periodic boundary condition is applied in the $x$-direction, while the open boundary condition with fifteen lattice sites is used in the $y$-direction. (b). The spatial profile of a chiral topological boundary state. (c). The phase diagram of the (2+1)-dimensional topological space-time crystal in the ($k_\delta^x$, $k_\delta^y$) space. Other parameters are the same to that used in (a). (d). The schematic diagram of the (2+1)-dimensional topological space-time circuit with $L$=5. The right-top inset

presents the structure of the time-varying coupling element, where the color map quantifies the position-dependent initial phase of external voltages. The right-bottom inset shows the values of various grounding INICs and resistors. (e). The photograph image of the fabricated circuit sample. The inset displays an enlarged view of a time-varying INIC along y-axis. (f1)-(f4). The measured spatial distributions of $V_{x,y}(t)$ at $t$=0, 3.7ms, 6.3ms, and 9.4ms.

**(3+1)-Dimensional Weyl-semimetal topolectrical space-time circuits.** In the last part, we propose and investigate the (3+1)-dimensional ($x$, $y$, $z$, $t$) topological space-time crystal exhibiting Weyl-semimetal physics. The corresponding Hamiltonian is expressed as

$$H_{3D}(x,y,z,t) = \sum_{x,y,z\in[1,L]}\{i[J_x(x,y,z,t)a^\dagger_{x,y,z}a_{x+1,y,z}$$
$$+J_y(x,y,z,t)a^\dagger_{x,y,z}a_{x,y+1,z} + J_z a^\dagger_{x,y,z}a_{x,y,z+1}] + H.c.\}. \quad (5)$$

Here, $J_z$ is the constant coupling along the $z$-axis. Time-varying coupling terms along the $x$- and $y$-axes are in the form of $J_x(x,y,z,t) = \Delta\cos[k_\delta^x(x+0.5) + k_\delta^y y + k_\delta^z z - \Omega t]$ and $J_y(x,y,z,t) = \Delta\sin[k_\delta^x x + k_\delta^y(y+0.5) + k_\delta^z z - \Omega t]$, which obey four discrete space-time translation symmetries as $J_i(x,y,z,t) = J_i(x,y,z,t-2\pi/\Omega)$, $J_i(x,y,z,t) = J_i(x+1,y,z,t+k_\delta^x/\Omega)$, $J_i(x,y,z,t) = J_i(x,y+1,z,t+k_\delta^y/\Omega)$ and $J_i(x,y,z,t) = J_i(x,y,z+1,t+k_\delta^z/\Omega)$ with $i \propto x,y$. Similar to the (1+1)- and (2+1)-dimensional topological space-time crystals, the (3+1)-dimensional space-time crystal can also be described by an energy-enlarged Floquet Hamiltonian $H(k_x,k_y,k_z)$ (see Supplementary Note 9). In this case, under the assumption with relatively weak time-varying couplings, where the energy overlap only exists between adjacent Floquet sectors, the low-energy physics around $\varepsilon = -0.5\Omega$ can be determined by the two-band effective Hamiltonian as (see Supplementary Note 9 for details)

$$H_{eff}(k_x,k_y,k_z) = \begin{bmatrix} 2J_z\sin(k_z)+0.5\Omega & \Delta\sin(k_x + 0.5k_\delta^x) - i\Delta\sin(k_y + 0.5k_\delta^y) \\ \Delta\sin(k_x + 0.5k_\delta^x) + i\Delta\sin(k_y + 0.5k_\delta^y) & 2J_z\sin(k_z + k_\delta^z) - 0.5\Omega \end{bmatrix}, \quad (6)$$

where the analytical expression for the quasi-energies are written as $E_\pm = J_z[\sin(k_z) + \sin(k_z + k_\delta^z)] \pm \sqrt{(\Delta\sin(k_x + 0.5k_\delta^x))^2 + (\Delta\sin(k_y + 0.5k_\delta^y))^2 + (J_z\sin(k_z) - J_z\sin(k_z + k_\delta^z)+0.5\Omega)^2}$. It is worth noting that the effective Hamiltonian $H_{eff}(k_x,k_y,k_z)$ takes the same form as the static Weyl Hamiltonian [71]. In this case, the space-time Weyl points are except to emerge in the quasi-momentum space when the following equations are satisfied: $\sin(k_x + 0.5k_\delta^x) = \sin(k_y + 0.5k_\delta^y) = 0$ and $J_z\sin(k_z) - J_z\sin(k_z + k_\delta^z)+0.5\Omega = 0$. Specifically, eight Weyl points at $E_\pm = 0$ (corresponds to $\varepsilon = -0.5\Omega$) can emerge in the quasi-momentum space $(k_x,k_y,k_z) = (\frac{2(n_x+1)\pi-k_\delta^x}{2},\frac{2(n_y+1)\pi-k_\delta^y}{2},\frac{2(n_z+1)\pi-k_\delta^z}{2})$ with $n_{x,y,z} = 0$ or 1, if we set $k_\delta^z = (\sqrt{3}-1)\pi$ and $J_z = -0.274\Omega$.

To further illustrate the existence of space-time Weyl points in the enlarged multi-band Floquet

Hamiltonian $H(k_x, k_y, k_z)$, we calculate its quasi-energy band structure in the $(k_x, k_y)$ plane for $k_z = 3.26$ and $k_z = 0.72$, as shown in Figure 3(a1). Other parameters are set as $\Delta = 0.5\Omega$, $J_z = -0.274\Omega$. $k_\delta^x = k_\delta^y = \frac{\sqrt{5}-1}{2}\pi$ and $k_\delta^z = (\sqrt{3}-1)\pi$. We observe that four space-time Weyl points with $\varepsilon = -0.42\Omega$ appear at $(k_x, k_y, k_z) = (0.691\pi + n_x\pi, 0.691\pi + n_y\pi, 3.26)$ with $n_{x,y} = 0$ or 1. Additionally, four other space-time Weyl points with $\varepsilon = -0.58\Omega$ are found at $(k_x, k_y, k_z) = (0.691\pi + n_x\pi, 0.691\pi + n_y\pi, 0.72)$ where $n_{x,y} = 0$ or 1. Notably, the $k_x$ and $k_y$ values of these Weyl points match those predicted by the low-energy effective Hamiltonian of Eq. (6), with slight deviations in $k_z$ and energy. These deviations arise from the influence of higher-energy bands. However, due to the robustness of the Weyl points, eight Weyl points still persist in quasi-momentum space. Furthermore, similar to the (1+1)- and (2+1)-dimensional cases, the space-time Weyl points in different Floquet sectors exhibit variations in both energy and momentum. For example, the space-time Weyl points with $\varepsilon = 0.58\Omega$ appear at $(k_x, k_y, k_z) = (0.691\pi + n_x\pi, 0.691\pi + n_y\pi, 3.26) + (k_\delta^x, k_\delta^y, k_\delta^z)$ with $n_{x,y} = 0$ or 1, and those with $\varepsilon = 0.42\Omega$ appear at $(k_x, k_y, k_z) = (0.691\pi + n_x\pi, 0.691\pi + n_y\pi, 0.72) + (k_\delta^x, k_\delta^y, k_\delta^z)$ with $n_{x,y} = 0$ or 1, as shown in Fig. 3(a2). This demonstrates the translations of space-time Weyl points in adjacent Floquet sectors as $\varepsilon \to \varepsilon + \Omega$ and $(k_x, k_y, k_z) \to (k_x + k_\delta^x, k_y + k_\delta^y, k_z + k_\delta^z)$.

Additionally, it is well known that the presence of Weyl points can induce the emergence of Weyl surface states in structures with open boundaries. To examine this, we calculate the dispersion of surface states for a supercell (see right inset of Fig. 3b), with periodic boundary conditions being applied in the $z$ and $v = \frac{1}{\sqrt{2}}(x + y)$ directions, while the $u = \frac{1}{\sqrt{2}}(x - y)$ direction is treated with open boundary condition with fifteen sites. The energy dispersion of surface states with $k_z = \pi$ is displayed in Fig. 3b, with the color map quantifies the strength of surface localization of all eigenmodes. It is shown that topological surface states emerge within the bandgap around $\varepsilon = \pm 0.5\Omega$. Moreover, a key manifestation of the topological nature of a Weyl system is the presence of topological surface states that form arcs connecting bulk states with distinct topological characteristics. To illustrate this, we calculate the Fermi arcs of the Weyl space-time crystal at $\varepsilon = -0.42\Omega$ in the $(k_v, k_z)$ space, as shown in Fig. 3c. The results clearly reveal two Weyl points at $k_z = 3.26$ (marked by the blue and red dots with opposite topological charges) and two closed circles of bulk states (marked by the blue and red lines), which originate from two $\varepsilon = -0.58\Omega$ Weyl points at $k_z = 0.72$. These Fermi arcs connect the Weyl points and bulk-state circles with opposite topological charges. To further clarify the quasi-energy spectrum of the (3+1)-dimensional Weyl topological space-time crystal, we calculate the eigen-spectrum of the finite (3+1)-dimensional space-time crystal (L=15) by cutting the *xy*-plane along $v$ and $u$ directions, as shown in Fig. 3d. It is shown that the Weyl points-induced surface states can appear within band gaps around $\varepsilon=\pm0.5\Omega$. The spatial profiles of a Weyl surface state at $\varepsilon = -0.497\Omega$, and a bulk state around the Weyl point at $\varepsilon = -0.41\Omega$ are plotted in Figs. 3(e1)-(e2), showing a strong boundary localization of the Weyl surface state and the property with an extended spatial profile of the bulk state. In addition, Weyl surface states of (3+1)-dimensional topological space-time crystals can also exhibit the chiral propagation behavior along the surface (See Supplementary Note 10 for simulation results).

In the following, we focus on the experimental implementation of our proposed (3+1)-dimensional Weyl topological space-time crystal. It is worth noting that the experimental realization of 3D non-equilibrium topological systems has never been reported up to now, due to the difficulty on the realization of 3D time-varying modulations. Here, we design a 3D topolectrical space-time circuit (as shown in Fig. 3f) to realize surface states of (3+1)-dimensional Weyl topological space-time crystal. The space-time modulated hoppings in *xy*-plane are achieved by time-varying INICs $\pm\frac{v_x(x,y,z,t)}{20CR_x}$ and $\pm\frac{v_y(x,y,z,t)}{20CR_y}$, where the initial phases of external voltages $v_x(x,y,z,t) = V_0\cos[\omega_0 t + \varphi_x(x,y,z)]$ and $v_y(x,y,z,t) = V_0\sin[\omega_0 t + \varphi_y(x,y,z)]$ take on the form of $\varphi_x(x,y,z) = k_\delta^x(x + 0.5) + k_\delta^y y + k_\delta^z z$ and $\varphi_y(x,y,z)=k_\delta^x x + k_\delta^y(y + 0.5) + k_\delta^z z$, as shown by the color map, respectively. The constant hopping along *z*-axis is realized by INICs $\pm\frac{1}{R_0}$, as illustrated in the right inset. Furthermore, all circuit nodes are connected to outputs of adjacent multipliers to ensure the time-invariant onsite potential. In this case, the eigen-equation of our designed 3D topolectrical space-time circuit takes on the same form with that of (3+1)-dimensional Weyl topological space-time crystal, where the effective tight-binding parameters are $\frac{v_x(x,y,z,t)}{20CR_x}/\omega_0 = J_x(x,y,z,t)/\Omega$, $\frac{v_y(x,y,z,t)}{20CR_y}/\omega_0 = J_y(x,y,z,t)/\Omega$, and $-\frac{1}{CR_0}/\omega_0 = J_z/\Omega$ (see Supplementary Note 11 for details). Here, we set circuit parameters as $C = 300nF$, $R_0 = 14.6k\Omega$, $R_x = 1k\Omega$, and $R_y = 1k\Omega$. The driving frequency and amplitude are equal to $\omega_0 = 0.833kHz$ and $V_0 = 2.5V$. Thus, the quasi-frequency of the circuit is related to the quasi-energy in Fig. 3a as $\frac{2\pi f}{\omega_0} = \frac{\varepsilon}{\Omega}$ with $\frac{V_0}{20CR_x}/\omega_0 = \frac{V_0}{20CR_y}/\omega_0 = \Delta/\Omega = 0.5$, and $\frac{1}{CR_0}/\omega_0 = J_z/\Omega = -0.2739$.

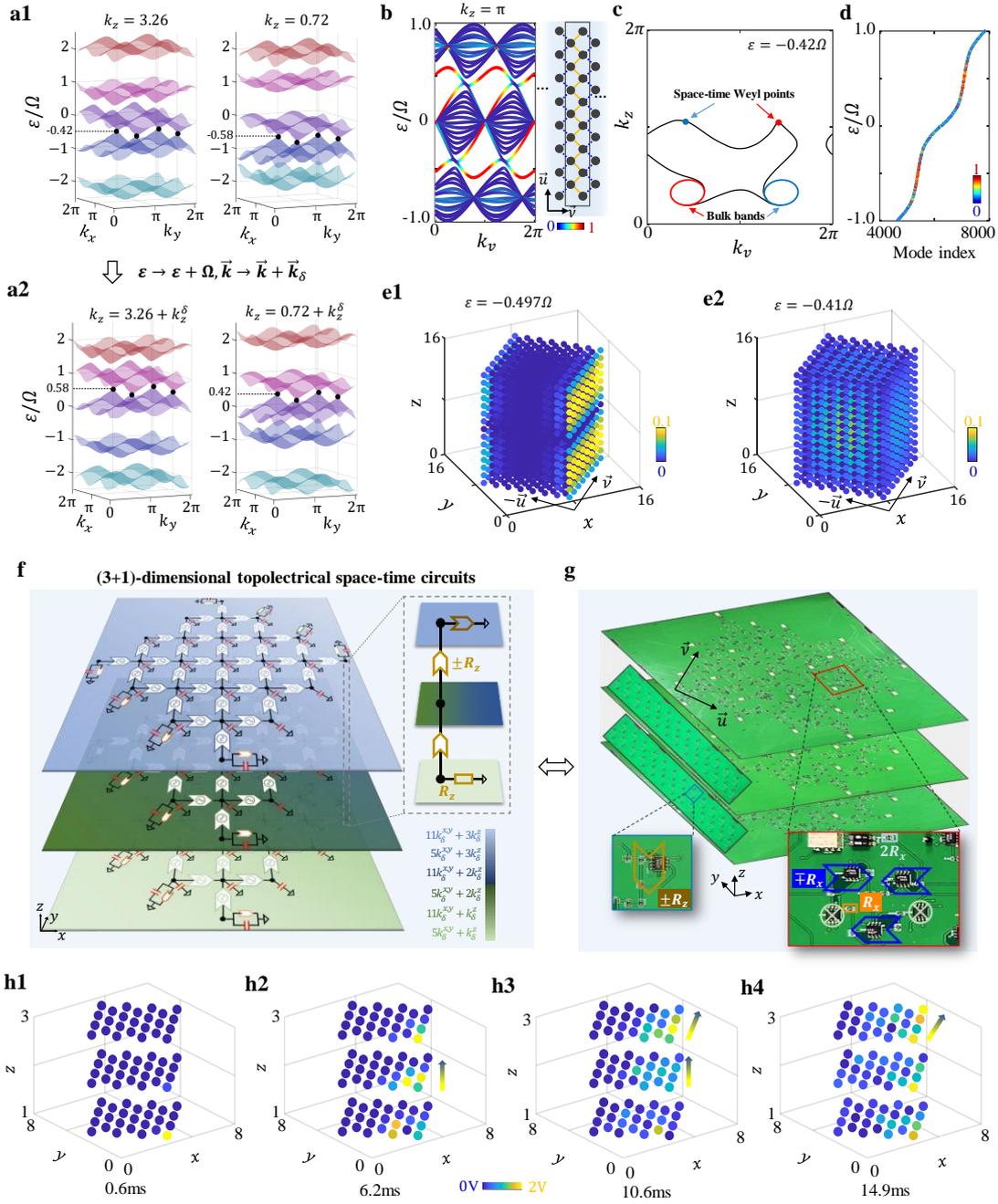

**Figure 3. Theoretical and experimental results of (3+1)-dimensional topolectrical space-time circuits.** The quasi-energy band structure of the enlarged Floquet Hamiltonian $H(k_x, k_y, k_z)$ in the $(k_x, k_y)$ plane for $k_z = 3.26$ and $k_z = 0.72$ in (a1) and $k_z = 3.26 + k_\delta^z$ and $k_z = 0.72 + k_\delta^z$ in (a2). The parameters are set as $\Delta = 0.5\Omega$, $J_z = -0.274\Omega$. $k_\delta^x = k_\delta^y = \frac{\sqrt{5}-1}{2}\pi$ and $k_\delta^z = (\sqrt{3}-1)\pi$. (b). The energy dispersion of surface states with $k_z = \pi$ in the (3+1)-dimensional Weyl topological space-time crystal, where periodic boundary conditions are applied in the $\vec{z}$ and $\vec{v}$ directions, while the $\vec{u}$-direction is treated with open boundary condition with fifteen lattice sites, as illustrated in the inset. (c). The result of Fermi arcs in the (3+1)-dimensional Weyl space-time crystal at $\varepsilon = -0.42\Omega$. (d). The quasi-energy spectrum of the finite (3+1)-dimensional Weyl topological space-time crystal, where the open boundary conditions are applied by cutting the $xy$-plane along $v$ and $u$ directions. (e1) and (e2). The spatial profile of a Weyl surface state and bulk state in the finite (3+1)-dimensional space-time crystal. (f). The schematic diagram of the (3+1)-dimensional topolectrical space-time circuit. The right inset presents the time-invariant coupling along the *z* direction. The color map manifests the position-dependent initial phase of external voltages. (g). The photograph image of the fabricated (3+1)-dimensional topolectrical space-time circuits. Two insets present the circuit realizations of couplings along *z*- and *x*-axis. (h1)-(h4). Measured spatial distributions of $V_{x,y,z}(t)$ at t=0.6ms, 6.2ms, 10.6ms and 14.9ms with the initial voltage at a corner node.

Then, we fabricate the designed (3+1)-dimensional topolectrical space-time circuit with seven units along *x*/*y*-axis (four units along $\vec{v}$ and $\vec{u}$ directions) and three units along *z*-axis, and the image of the circuit sample is presented in Fig. 3g. Two insets present the circuit realizations of time-invariant and time-varying couplings along *z*- and *x*-axes, respectively. To detect the Weyl space-time surface state, we measure the temporal evolution of voltage signals at all circuit nodes in the Weyl space-time circuit with the initial voltage being 5V at the right-bottom corner node and 0V for other circuit nodes. The spatial distributions of measured voltages $|V_{x,y,z}(t)|$ at *t*=0.6ms, 6.2ms, 10.6ms and 14.9ms are plotted in Figs. 3(h1)-3(h4). It is clearly shown that the input voltage on the right-bottom corner primarily propagates along surface nodes in the $\vec{v} = \frac{1}{\sqrt{2}}(\vec{x} + \vec{y})$ direction, and the extremely weak voltage signals transport along the $\vec{u} = \frac{1}{\sqrt{2}}(\vec{x} - \vec{y})$ axis. Such a chiral propagation of Weyl surface states is consistent with simulations (see Supplementary Note 12). Thus, the above experimental results clearly demonstrate the achievement of Weyl surface states in (3+1)-dimensional topolectrical space-time circuits.

**Conclusion.** In conclusion, we have reported the first experimental implementation of topological space-time crystals by constructing topolectrical space-time circuits. We have directly observed midgap topological edge modes protected by the generalized particle-hole symmetry in (1+1)-dimensional topolectrical space-time circuits. In addition, by constructing time-modulated topolectrical circuits that sustain (2+1)-dimensional discrete space-time translation symmetries, we have experimentally realized topological space-time chiral edge states. The chiral behavior of topological boundary propagation has been observed through voltage dynamics. Finally, we have not only theoretically proposed the (3+1)-dimensional space-time Weyl semimetals, but also experimentally demonstrated Weyl surface states induced by space-time Weyl points using three-dimensional topolectrical space-time circuits.

It is worth noting that our PCB-integrated topolectrical space-time circuits can also be extended to CMOS chips operating in the microwave region, with a wide range of potential applications anticipated in the fields of 5G wireless systems and radar technology in the future [14, 65]. Furthermore, other artificial platforms capable of effective temporal modulations, such as photonic waveguides with the propagation direction acting as the time dimension [12], ultra-cold bosonic atoms in optical superlattice [72], time-modulated acoustic cavities [73] and light propagation in coupled optical fiber [74], are anticipated to enable the engineering of classical and quantum wave systems integrated with unique space-time topology. Our work establishes a foundation for exploring intricate space-time topological matters and holds great potential as a crucial component for experimental simulations involving diverse time-modulated Hamiltonians with extraordinary dynamical properties.

**Methods**

**Sample fabrications.** We exploit electric circuits by using LCEDA program software, where the PCB composition, stack-up layout, internal layer and grounding design are suitably engineered. The designed PCBs have six layers. Except for the top and bottom two layers, there are two power layers and two grounding layers. For the realization of the INIC, we use the operational amplifier (OpAmp) of LT1363. In addition, two surface-mounted resistors are used as the auxiliary resistors in the positive and negative feedback loops of the OpAmp. The OpAmps and multiplier (AD633JNZs) are supplied by external voltages of $\pm 15V$. Besides, the values of all circuit elements are large enough to ignore the influence of effective resistances and parasitic capacitances in the circuit sample.

**Experimental details.** As for the time-domain voltage measurement, we initially establish the system's initial state by assigning an initial voltage to each lattice node. In other words, at the initial moment, each lattice node is connected to a DC voltage source or the ground (0V). In this case, we use the relay model and a DIP switch to connect all circuit nodes with respect to corresponding DC voltage sources or GND layer. Specifically, the relay model is controlled by a voltage signal of 5V through the switch. With this setting, the required initial voltages can be applied to all circuit nodes at the same time. In addition, it is worth noting that the external driving voltages, which are injected into input ports of multipliers, are also needed to be calibrated with respect to the initial voltage on circuit nodes. Hence, for this purpose, we use a few amounts of signal generators (FY2300-12M) to create voltage signals with the same frequency but different initial phases (matched to the requirement of position-dependent initial phase in topological space-time crystal). Then, the external driving voltages are also controlled by the same mechanical switch used for setting the initial state. As for the measurement of voltage dynamics, we connect circuit nodes to 4-channel oscilloscope DSO7104B (Agilent Technologies) by coaxial cables and measure the voltage signal after the switch is turned off.

**Data Availability.** All data are displayed in the main text and Supplementary Information.

**Acknowledgements.** This work is supported by the National Key R & D Program of China under Grant No. 2022YFA1404900, National Science Foundation of China No. 12422411, Young Elite Scientists Sponsorship Program by CAST No. 2023QNRC001 and Beijing Natural Science Foundation No. 1242027.

**Author contributions statement.** W. Zhang finished the theoretical scheme and designed topolectrical space-time circuits with the help of W. Cao and L. Qian. W. Cao and L. Qian performed the experiments with the help of H. Yuan. W. Zhang and X. Zhang wrote the manuscript. X. Zhang initiated and designed this research project.

**Competing interests statement.** The authors declare no competing interests.